\documentclass[aps,prl,twocolumn,groupedaddress,showpacs]{revtex4}

\usepackage{graphicx}

\begin{document}

\title{Tomographic RF Spectroscopy of a Trapped Fermi Gas at Unitarity}

 \author{Y. Shin} \email{yishin@mit.edu}
\author{C.~H. Schunck}
\author{A. Schirotzek}
\author{W. Ketterle}

\affiliation{Department of Physics, MIT-Harvard Center for
Ultracold Atoms, and Research Laboratory of Electronics,
Massachusetts Institute of Technology, Cambridge, Massachusetts,
02139}

\date{\today}

\begin{abstract}
We present spatially resolved radio-frequency spectroscopy of a
trapped Fermi gas with resonant interactions and observe a
spectral gap at low temperatures. The spatial distribution of the
spectral response of the trapped gas is obtained using in situ
phase-contrast imaging and 3D image reconstruction. At the lowest
temperature, the homogeneous rf spectrum shows an asymmetric
excitation line shape with a peak at $0.48(4) \varepsilon_F$ with
respect to the free atomic line, where $\varepsilon_F$ is the
local Fermi energy.\end{abstract}


\maketitle

Ultracold Fermi gases near a Feshbach resonance provide new
insight into fermionic superfluidity, allowing the study of the
crossover from Bardeen-Cooper-Schrieffer (BCS) superfluids of
Cooper pairs to Bose-Einstein condensates (BECs) of dimer
molecules. Many aspects of the BEC-BCS crossover, including
superfluidity~\cite{ZAS05}, have been experimentally
investigated~\cite{Grimm07}. The properties of Fermi gases on
resonance at unitarity, where the amplitude of the scattering
length between the fermion atoms diverges to infinity and the
system shows universal behavior~\cite{Ho04}, are of great
importance to understand the crossover physics. Precision
measurements of  the critical temperature~\cite{KTT05}, the
interaction energy~\cite{SGR06} and collective
excitations~\cite{ARK07} have presented stringent quantitative
test to the theoretical description of strongly interacting Fermi
gases.

Radio-frequency (rf) spectroscopy measures an excitation spectrum
by inducing transitions to different hyperfine spin states. This
method has been employed in strongly interacting Fermi gases,
leading to the observation of unitarity limited
interactions~\cite{RJ03,GHZ03}, molecule formation on the BEC side
of the Feshbach resonance~\cite{RTB03} as well as pairing in the
crossover regime~\cite{CBA04,SSS07}. Rf spectroscopy provides
valuable information on the nature of the pairs. Since an rf
photon can dissociate bound molecules or fermion pairs into the
free atom continuum, the binding energy of the pairs or the
excitation gap is determined. Furthermore the excitation line
shape is related to the wave function of the pairs, e.g. larger
pairs have narrower lines. However, currently all experimental
measurements on the excitation spectrum in strongly interacting
Fermi gases~\cite{CBA04,GRJ05} have been performed with samples
confined in a harmonic trapping potential so that the spectral
line shape is broadened due to the inhomogeneous density
distribution of the trapped samples, preventing a more stringent
comparison with theoretical
predictions~\cite{KRT04,HCL05,OG05,YB06}.

In this Letter, we demonstrate spatially resolved rf spectroscopy
of a trapped, population-balanced Fermi gas at unitarity at very
low temperature. The spatial distribution of the rf-induced
excited region in the trapped gas was recorded with in situ
phase-contrast imaging~\cite{SZS06} and the local rf spectra were
compiled after 3D image reconstruction. In contrast to the
inhomogeneous rf spectrum, the homogeneous local rf spectrum
shows a clear spectral gap with an asymmetric line shape. We
observe that the spectral peak shifts by $0.48(4) \varepsilon_F$
to higher energy and that the spectral gap is $0.30(8)
\varepsilon_F$ with respect to the free atomic reference line,
where $\varepsilon_F$ is the local Fermi energy. This new
spectroscopic method overcomes the line broadening problem for
inhomogeneous samples and provides homogeneous rf spectra of a
resonantly interacting Fermi gas revealing the microscopic
physics of fermion pairs.

We prepared a degenerate Fermi gas of spin-polarized $^6$Li atoms
in an optical trap, using laser cooling and sympathetic cooling
with $^{23}$Na atoms, as described in Ref.~\cite{HGS03}. An equal
mixture of the two lowest hyperfine states $|1\rangle$ and
$|2\rangle$ (corresponding to the $|F=1/2,m_F=1/2\rangle$ and
$|F=1/2,m_F=-1/2\rangle$ states at low magnetic field) was created
at a magnetic field $B=885$~G. A broad Feshbach resonance located
at $B_0=834$~G strongly enhanced the interactions between the two
states. The final evaporative cooling by lowering the trap depth
and all spectroscopic measurements were performed at $B=833$~G.
The total atom number was $N_t = 1.0 \pm 0.1 \times 10^7$ and the
radial (axial) trap frequency was $f_r=129$~Hz ($f_z=23$~Hz). The
Fermi energy (temperature) of a noninteracting equal mixture with
the same total atom number is $E_{F}=h (f_r^2 f_z)^{1/3} (3
N_t)^{1/3} = h \times 22.3$~kHz ($T_{F}=E_{F}/k_B =1.07~\mu$K),
where $h$ is Planck's constant and $k_B$ is Boltzmann's constant.
The ratio of the sample temperature $T$ to $T_F$ of $\approx
0.06$ was determined by fitting a finite temperature Thomas-Fermi
(TF) distribution to the whole cloud after expansion.

Rf spectroscopy was performed by driving atoms in state
$|2\rangle$ to the next lowest hyperfine state $|3\rangle$
(corresponding to $|F=3/2, m_F=-3/2\rangle$ at low field) that
was initially empty. After applying  an rf pulse of 1~ms, we
directly measured the in situ distribution of the density
difference $n_d\equiv n_1 - n_2$ in the excited sample, using a
phase-contrast imaging technique~\cite{SZS06}, where $n_1$ and
$n_2$ are the densities of atoms in the states $|1\rangle$ and
$|2\rangle$, respectively. The frequency of the imaging beam was
set to make the net phase shift in the sample proportional to the
density difference $n_d$, which was done by zeroing the optical
signal in an equal mixture without applying an rf
pulse~\cite{SZS06}. Since the initial atom densities in state
$|1\rangle$ and $|2\rangle$ are equal, the density difference
$n_d$ represents the atom number depletion in state $|2\rangle$,
the spectral response $I$~\cite{footnote1}.

\begin{figure}
\begin{center}
\includegraphics[width=3.2in]{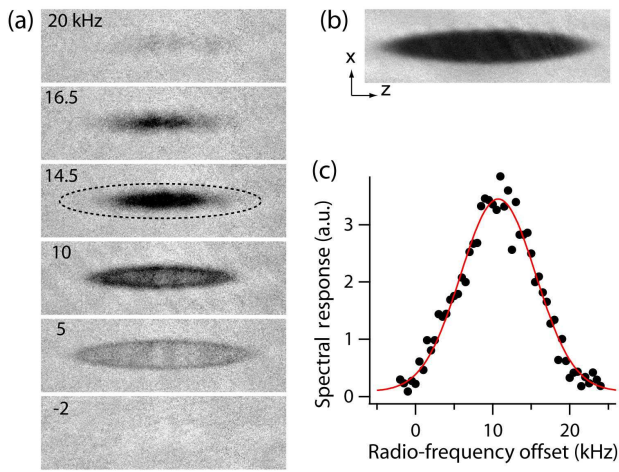}
\caption{(color online) Radio-frequency (rf) spectroscopy of a
Fermi gas with in situ phase-contrast imaging. (a) After applying
an rf pulse, the spatial distribution of the density difference
between state $|1\rangle$ and $|2\rangle$ is recorded with the
phase-contrast imaging technique~\cite{SZS06}. The density
depletion reflects the spin excitation induced by the rf pulse.
The dashed line indicates the size of the trapped sample. (b) An
absorption image of an equal mixture without applying an rf
pulse. The field of view for each image is
$205~\mu$m$~\times~680~\mu$m. (c) Overall rf spectrum of the
inhomogeneous sample is obtained by integrating the signal in the
phase-contrast images. The red line is a Gaussian fit to the
spectrum.\label{f:imaging}}
\end{center}
\end{figure}

The total spectral response, obtained by integrating over the
phase contrast images, reproduces earlier
results~\cite{CBA04,SSS07}. The phase-contrast images now reveal
the nature of the observed line shape (Fig.~\ref{f:imaging}). The
spectral response strongly depends on position. The inner region
of the cloud, which is at higher density, shows a higher
resonance frequency. The integrated inhomogeneous spectrum peaks
at the rf offset $\Delta \nu_i \approx 10$~kHz~\cite{footnote2}.
The spatially resolved images reveal that at this frequency, no
excitations occur in the center of the cloud, but rather in a
spatial shell. The rf offset $\Delta \nu$ is measured with
respect to the resonance frequency of the $|2\rangle - |3\rangle$
transition in the absence of atoms in state
$|1\rangle$~\cite{footnote3}.

Local rf spectra $I(r,\Delta\nu)$ are compiled from the
reconstructed 3D radial profiles of the density difference. A
phase-contrast image contains the 2D distribution of the column
density difference integrated along the imaging line,
$\tilde{n}_d(x,z)\equiv \int n_d(\vec{r})~dy$. The excited regions
have an elliptical shape with the same aspect ratio as the trap,
$\lambda= f_z/f_r$, confirming the validity of the local density
approximation. Therefore, we can use elliptically averaged
profiles of the column density difference, $\tilde{n}_d(r)$, to
improve the signal-to-noise ratio, where the ellipse for
averaging is defined as $x^2+\lambda^2 z^2=r^2$. The 3D radial
profile $n_d(r)$ is calculated using the inverse Abel
transformation of $\tilde{n}_d(r)$~\cite{Bra86} and
$I(r,\Delta\nu) \propto n_d(r;\Delta\nu)$.

\begin{figure}
\begin{center}
\includegraphics[width=3.0in]{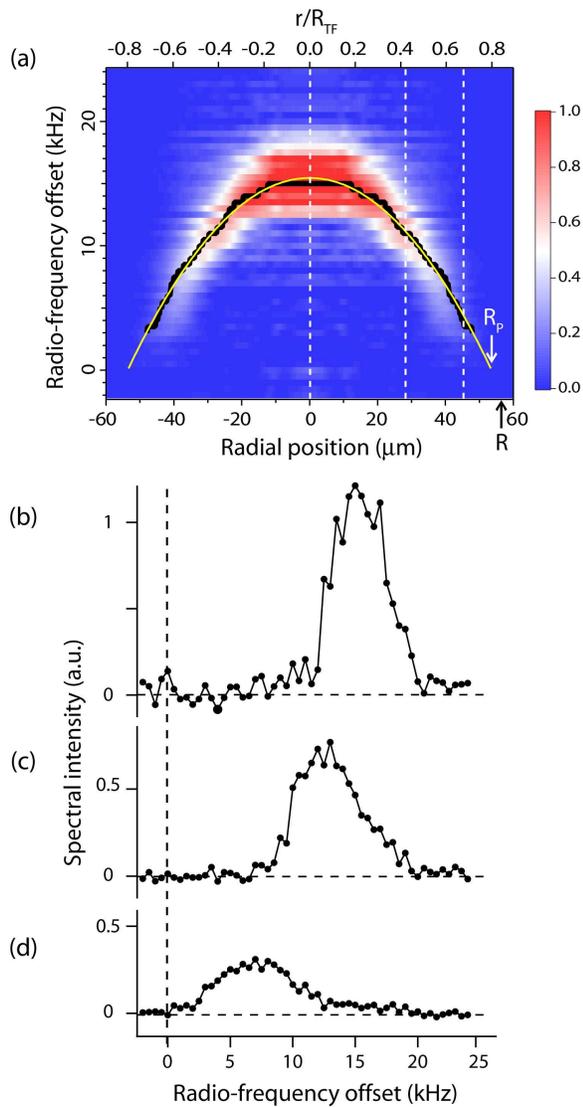}
\caption{(color online) Spatially resolved rf spectroscopy of a
trapped Fermi gas. (a) The spectral intensity $I(r,\Delta\nu)$ was
obtained from the reconstructed 3D profiles of the density
difference. See text for the description of the reconstruction
method. Local rf spectra are shown for (b) $r=0~\mu m$, (c)
$r=24~\mu m$, and (d) $r=40~\mu m$, whose positions are marked by
vertical dashed lines in (a). Each spectrum is obtained by
spatially averaging over $2.5~\mu$m. $R_{TF}$ is the radial
Thomas-Fermi radius for a noninteracting Fermi gas with the same
atom number. The spectral peak position $\Delta\nu_p(r)$ in the
local rf spectra is marked by the black line in (a). The
determination of $\Delta\nu_p$ is limited to $r<48~\mu$m due to
the signal-to-noise ratio. The yellow line is a parabolic fit to
$\Delta\nu_p(r)$. The radius determined by extrapolating the fit
to zero rf offset is $R_p=53.6~\mu$m indicated by the white down
arrow. The black up arrow indicates the radius of the trapped
sample, $R=56.6~\mu$m, measured independently from absorption
images like Fig.~\ref{f:imaging}(b).\label{f:2dspectrum}}
\end{center}
\end{figure}

With this technique, we obtain homogeneous rf spectra as a
function of the 3D position, shown in Fig.~\ref{f:2dspectrum}.
These spectra are the main result of this paper and we now
discuss their features and implications for our system. The local
homogeneous rf spectra shows a spectral gap. The spectral peak is
shifted away from the atomic reference line by much more than its
line width. Such a gap is not observed in the inhomogeneous rf
spectrum (Fig.~\ref{f:imaging}(c)) where the Gaussian wings
overlap with the position of the free atomic line. Furthermore,
the local rf spectrum reveals an asymmetric line shape of the
excitation spectrum. For the central region, the peak is located
at $\Delta\nu_p \approx 15$~kHz and the spectral gap, defined as
the minimum energy offset for excitation, is $h \Delta \nu_g
\approx h \times 10$~kHz.

The spectral peak position $\Delta \nu_p$ in the local rf spectra
shows a parabolic dependence on the radial position
(Fig.~\ref{f:2dspectrum}(a)). This can be explained by unitarity,
which demands that all energetic quantities scale with the Fermi
energy. At unitarity, the only relevant energy scale in the
system is the Fermi energy $\varepsilon_F \equiv \hbar^2 (6 \pi^2
n)^{2/3} / 2 m$~\cite{HEI01,Ho04}, where $n$ is the atom density
in one spin state and $m$ is the atomic mass, so the energetic
quantities such as chemical potential $\mu$ and pairing gap
energy $\Delta$ are proportional to $\varepsilon_F$, i.e.,
$\mu=\xi \varepsilon_F$ and $\Delta=\eta \varepsilon_F$ with the
universal parameters $\xi$ and $\eta$. Therefore the excitation
spectrum should also scale with the Fermi energy. In an external
harmonic potential $V(r) \propto r^2$, the local Fermi energy
$\varepsilon_F(r)=\mu(r)/\xi=(\mu_0-V(r))/\xi=\varepsilon_{F0}(1-
r^2/R^2)$, where $\mu_0$ is the global chemical potential,
$\varepsilon_{F0}$ is the local Fermi energy at the center, $R$
is the radius of the trapped sample and $\varepsilon_{F0}=\mu_0
/\xi = V(R)/\xi$. The spectral peak position $\Delta \nu_p(r)$
reflects the parabolic radial dependence of the local Fermi
energy $\varepsilon_F(r)$.

The local Fermi energy at the center is determined from
$\varepsilon_{F0}= (R/R_{TF})^{-2} E_F$, where $R_{TF}$ is the
radial Thomas-Fermi radius for a noninteracting Fermi gas with
the same atom number. We obtain $R_{TF}=67.3 \pm 1.1~\mu$m for
the measured total atom number and trap frequencies. The radius
of the trapped sample was measured to be $R=56.6 \pm 1.8~\mu$m,
using absorption images like in Fig.~\ref{f:imaging}(b) and
fitting the non-saturated outer wing profile to a
zero-temperature TF distribution, giving
$\varepsilon_{F0}=h\times 31.5 \pm 2.5$~kHz. We estimate the
universal parameter $\xi = (R/R_{TF})^4 = 0.50 \pm 0.07$, which is
in good agreement with previous
measurements~\cite{OHG02,GHG03,BAR04b,BKC04,KTT05,PLK06,SGR06} and
Quantum Monte Carlo calculations~\cite{CCP03,ABC04,CR05} ($\xi
\equiv 1+\beta$ in some references).

The local spectrum at the center of our sample shows the spectral
peak at $h\Delta\nu_p = 0.48(4)\varepsilon_{F0}$ and the spectral
gap of $h\Delta\nu_g = 0.30(8)\varepsilon_{F0}$. We determine a
radius $R_p$ such that $\Delta\nu_p(R_p)=0$, extrapolating the
parabolic curve fit of $\Delta\nu_p$ to zero rf offset
(Fig~\ref{f:2dspectrum}(a)). $R_{p}=53.6~\mu$m is slightly
smaller than the measured radius $R$, which we attribute to
finite temperature effects. Previous studies of rf spectroscopy
of Fermi gases~\cite{CBA04,SSS07} demonstrated that the spectral
peak shifts to higher energy at lower temperature, which is
interpreted as the increase of the pairing gap energy. In the
outer region of lower density, the local $T/T_F$ becomes higher,
consequently reducing $h\Delta\nu_p / \varepsilon_F$. The
observation of $R_p$ being close to $R$ implies that our
experiment is very close to the zero temperature limit. From the
relation $T/T_F(r) \propto (1-r^2/R^2)^{-1}$, we can estimate
$T/T_F(R_p)\approx 15 \times T/T_F(0)$. If we assume that the
pairing gap energy starts emerging at $T/T_F \approx
0.6$~\cite{HCL05}, we might infer the local $T/T_F <0.05$ at the
center, close to our measured temperature. Although
$h\Delta\nu_p/\varepsilon_F$ is almost constant over the whole
sample, the line width increases in the outer region.

The homogeneous rf spectra measured in our experiment allow a
direct comparison with theoretical predictions. However, a
comprehensive theoretical interpretation of the rf spectrum of a
strongly interacting Fermi gas is not available yet and we
discuss here an extrapolation of BCS theory to strong unitarity
limited interactions. Rf spectroscopy measures a single-particle
spin excitation spectrum, since an rf photon changes the spin
state while imparting negligible momentum. The conventional
picture of rf spectroscopy of pairs is a photodissociation
process: the initial $|1\rangle-|2\rangle$ bound state, which can
be molecules or fermion pairs, breaks into free particles in
state $|1\rangle$ and $|3\rangle$. In a BCS superfluid, the free
particle in state $|1\rangle$ is regarded as a quasiparticle, so
after the spin transition the whole system can be described as
the excited BCS state with one quasiparticle and one free
particle in state $|3\rangle$. With the assumption of no
interactions between state $|1\rangle$ and $|3\rangle$, the rf
photon energy offset would be $h \Delta\nu = E_{-k} - \mu +
\hbar^2 k^2 /2m$, where the first term $E_{-k}=\sqrt{\Delta^2
+(\hbar^2 k^2 /2m - \mu)^2}$ is the energy cost for generating
one quasiparticle excitation with momentum $-k$, the second term
is for removing one atom in state $|2\rangle$, and the last term
is the kinetic energy of the atom in state $|3\rangle$ with
momentum $k$.

The measured FWHM line width is about two times narrower than
predicted by the simple model described above. The model spectrum
shows a very long tail corresponding to high momentum
contributions. This discrepancy might be due to modification of
the BCS expressions in the unitarity regime. The narrow peak
might imply that the pair wave function is narrower in momentum
space and therefore more spatially extended than the BCS
prediction. Another extension of our simple model should address
the interactions between atoms in state $|1\rangle$ and
$|3\rangle$. The mean-field interaction energies due to
$|1\rangle -|2\rangle$ and $|1\rangle -|3\rangle$ interactions
have been empirically assumed to have the same unitarity limited
value because of the proximity of a $|1\rangle - |3\rangle$
Feshbach resonance at B=690~G~\cite{CBA04,KRT04}. The recent
experiments with imbalanced mixtures and higher densities showed
some deviations from this assumption~\cite{SSS07}.

A localized spin excitation, induced by an rf pulse, eventually
diffuses over the sample. This ultimately limits the pulse
duration and therefore the spatial resolution. Using
phase-contrast imaging, we monitored the evolution of the spatial
structure in the excited sample with various delay time after
applying an rf pulse. The shell structure was well preserved even
after 5~ms and only some broadening was observed, showing that
during the 1-ms pulse the dynamic evolution of the density
difference profiles is negligible.

We found that atoms in state $|3\rangle$ rapidly decayed in the
sample, although the total atom number difference between state
$|1\rangle$ and $|2\rangle$ did not change over time. At the
center of the sample where the total atom density is about $8
\times 10^{12}~$cm$^{-3}$, the life time of atoms in state
$|3\rangle$ was measured to be less than 0.4~ms. Since a
$|1\rangle-|3\rangle$ mixture and a $|2\rangle-|3\rangle$ mixture
are stable at $B=833~$G~\cite{RGP}, the decay should be
associated with exoergic molecule formation via three-body
collisions involving one atom from each spin state. We observed
that the loss of one atom in state $|3\rangle$ was accompanied by
loss of one atom in state $|1\rangle$ and one atom in state
$|2\rangle$, supporting the three-body loss
mechanism~\cite{footnote4}.

Our new technique of spatially resolved rf spectroscopy should be
able to address the important questions also at finite
temperature. One question is whether the observed double peak
structure~\cite{CBA04,SSS07} of an atomic line and a pairing peak
is purely inhomogeneous, or whether it is possible to have local
coexistence of paired and unpaired atoms. This is of course
possible on the BEC side of the Feshbach resonance where in a
certain temperature range, bound molecules and thermally
dissociated free atoms coexist, but it is an open question, how
this picture will change in the BEC-BCS crossover.

In conclusion, we present spatially resolved rf spectroscopy of a
trapped Fermi gas, using the in situ phase-contrast imaging
technique. The homogeneous rf spectra of a Fermi gas at unitarity
provide a benchmark for a complete theoretical description, which
should reveal microscopic details of the paired states.

We thank M.~Zwierlein and R.~Grimm for stimulating discussions and
T.~Pasquini for critical reading of the manuscript. This work was
supported by the NSF and ONR.


\begin{thebibliography}{29}
\bibitem{footnote1}
Note that also atoms in state $|3\rangle$ contribute to the
optical signal. However, since their density is proportional to
$n_d$, the signal is still proportional to $n_d$.
\bibitem{footnote2}
The measured $h\Delta\nu_i /E_F \approx 0.45$
is higher than the values reported in Ref.~\cite{CBA04,SSS07},
which we attribute to lower temperature and less heating over the
short pulse duration.
\bibitem{footnote3}
The magnetic field was calibrated by determining the resonance
frequency $\nu_0$ of the $|2\rangle-|3\rangle$ transition in the
absence of atoms in state $|1\rangle$. $\nu_0=$ 81.700(1)~MHz
corresponds to a magnetic field of about 833.0(2)~G.
\bibitem{footnote4}
In our experiment, the maximum fractional atom loss in state
$|2\rangle$ ($|1\rangle$) was about 30\% (15\%) at $\Delta\nu
\approx 10~$kHz, which could not be reduced further due to the
signal-to-noise ratio. Including the spatial width of the
excitation region, we estimate that the local density depletion
was about 45\%.
\end{thebibliography}
\end{document}